\documentclass[pra,twocolumn,showpacs,superscriptaddress,aps]{revtex4-1}

\usepackage{amsmath}
\usepackage{amssymb}
\usepackage{graphicx}
\usepackage{placeins}

\begin{document}

\title{Molecular dissociation in few-cycle laser pulses:\\From attosecond to femtosecond pulse duration}

\author{T.~Fiedlschuster}	
\affiliation{Institut f\"{u}r Theoretische Physik, Technische Universit\"{a}t Dresden, D-01062 Dresden, Germany}
\affiliation{Max-Planck-Institut f\"ur Mikrostrukturphysik, Weinberg 2, D-06120 Halle, Germany}

\author{J.~Handt}
\affiliation{Institut f\"{u}r Theoretische Physik, Technische Universit\"{a}t Dresden, D-01062 Dresden, Germany}

\author{E.~K.~U.~Gross}	
\affiliation{Max-Planck-Institut f\"ur Mikrostrukturphysik, Weinberg 2, D-06120 Halle, Germany}
\affiliation{Fritz Haber Center for Molecular Dynamics, Institute of Chemistry, The Hebrew University of Jerusalem, Jerusalem 91904, Israel}

\author{R.~Schmidt}
\email{Ruediger.Schmidt@tu-dresden.de}
\affiliation{Institut f\"{u}r Theoretische Physik, Technische Universit\"{a}t Dresden, D-01062 Dresden, Germany}

\date{\today}

\begin{abstract}
The dissociation dynamics of diatomic molecules interacting with (near) optical laser pulses of different duration is investigated by an elaborate discussion of the electric field of the laser and by a direct solution of the time-dependent molecular Schr\"odinger equation. The systematic variation of the pulse duration from the electronic time scale (attoseconds) to the nuclear time scale (femtoseconds) shows that the employed few-cycle laser pulses lead to well-known but quite different dissociation mechanisms. A comparative calculation with a model pulse and  an experimentally realized attosecond pulse emphasizes that, and to what extent, a realistic modeling of the electric field is of central importance in the attosecond regime.           
\end{abstract}

\maketitle

\section{Introduction}\label{sec:intro}

Laser-induced dissociation of molecules (i.e. fragmentation without ionization) represents an ideal tool to probe the coupled dynamics of electrons and nuclei in atomic many-body systems within a wide range of time scales.

Today, laser systems are available which provide intense lasers with attosecond pulse duration ($1$ as $=10^{-18}$ s), corresponding to the electronic time scale. Hence, during the last years, large progress has been made to observe and even control the electronic dynamics, and, in molecules, its coupling to the nuclei (for reviews see e.g. \cite{krausz_2009,nisoli_2017}). Somewhat surprisingly, relatively little has been reported about the dissociation of diatomic molecules exposed to attosecond laser pulses. Indeed, the expected mechanism is seemingly known since the beginning of photochemistry or photodissociation physics (see Ref. \cite{schinke_1993} and references therein):
 
A single photon excites the electronic system suddenly and hence lifts the nuclei vertically to an excited repulsive potential energy surface (PES). Vertically means that this excitation is practically instantaneous on the nuclear time scale, such that the nuclei are assumed to be static during the interaction with the laser pulse. Afterwards, the repulsive nuclear forces lead to molecular dissociation on the inherently field-free Born-Oppenheimer (BO) PES. We will term this mechanism  "vertically excited dissociation" (VED).

On the other hand, the dissociation of diatomic molecules in intense lasers with femtosecond pulse duration ($1$ fs $=10^{-15}$ s), corresponding to the nuclear time scale, has been widely studied and a huge variety of phenomena have been found during the last decades (for a recent review see \cite{ibrahim_2017}). The most essential one represents the bond softening dissociation (BSD) mechanism \cite{bucksbaum_1990}. Other related effects are bond hardening \cite{giusti-suzor_1992}, multi-photon dissociation \cite{giusti-suzor_1990}, and zero-photon dissociation \cite{giusti-suzor_1992,posthumus_2000}, dynamical alignment \cite{charron_1994,uhlmann_2005} and anti-alignment \cite{frasinski_2001}, and centrifugal fragmentation \cite{fischer_2011}. The qualitative interpretation of all these phenomena is based on the Floquet picture \cite{shirley_1965,sambe_1973,bandrauk_1981,guerin_1997,chu_2004,fiedlschuster_2016}. Particularly the bond softening mechanism facilitates a physical transparent grasp of the dissociation dynamics.

\begin{figure}[htbp]
\begin{center}
\includegraphics[scale=0.5]{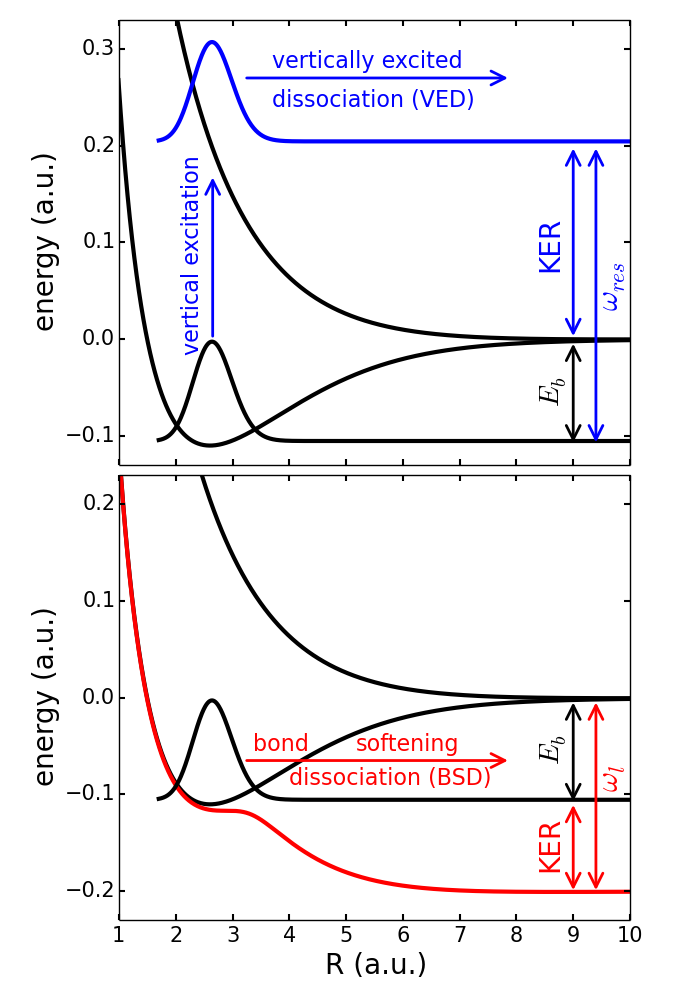}
\caption{Illustration of VED (upper panel) and BSD (lower panel) to aid the discussion in the text. In both panels, the black curves represent the ground state and first excited state BO PES and the initial nuclear wave packet in the BO ground state. The black vertical arrow indicates the binding energy $E_b$ of the molecule. The red curve shows the bond softening Floquet PES calculated for a continuous wave laser with carrier frequency $\omega_l=0.2$ a.u. and laser intensity $I=5\times10^{13}$ W/cm${}^2$.} 
\label{fig:fcp_bs}
\end{center}
\end{figure}

Both dissociation mechanisms, VED and BSD, are illustrated in Fig.~\ref{fig:fcp_bs} for a model system of $\textrm H_2^{\;+}$\!. Details of the model, for the present discussion still unimportant, will be outlined in Sec.~\ref{sec:model}. The two mechanisms can be elucidated as follows:

In the first case (VED, see upper panel of Fig.~\ref{fig:fcp_bs}), the extreme short laser suddenly excites the electron, leading to a vertical transition of the nuclear wave packet to the first excited BO-PES (blue vertical arrow and blue wave packet).  Afterwards, the nuclear wave packet evolves field-free on this PES, leading to dissociation (blue horizontal arrow). Several transitions can contribute to this excitation, if the necessary resonance frequencies $\omega_{res}$, given by the differences of the energies of the involved quantum states, are contained in the laser spectrum. 

In the second case (BSD, see lower panel of Fig.~\ref{fig:fcp_bs}), the femtosecond laser pulse and the molecule temporarily form an united quantum system exhibiting its own Hilbert space (see, e.g., \cite{guerin_1997} for details). The field-free BO states and corresponding BO PES, determining the field-free nuclear motion, are replaced by Floquet states and  Floquet PES, determining the laser-driven nuclear motion. For lucidity, we have plotted only the Floquet PES most relevant for BSD in Fig.~\ref{fig:fcp_bs}. This PES is qualitatively comparable to the ground state BO PES at small distances, $R< 2.5$ a.u., and to the first excited BO PES shifted down by the energy of one photon $\omega_l$ at large distances, $R>4$ a.u. The smooth behavior in the decisive transition region allows for nuclear motion towards  larger bond lengths, and finally dissociation (for a time-dependent visualization see "scenario1.mov" in the supplementary material to \cite{fiedlschuster_2017}).

In both cases, VED and BSD, the difference between $\omega_{res}$ and the binding energy $E_b$, respectively $\omega_l$ and $E_b$, is a measure of the mean value of the expected kinetic energy release (KER) of the dissociating fragments.

This short introduction naturally rises several questions. First of all, to what extend does the VED mechanism really apply for very short, but still finite strong laser pulses?  On the other side, what is the quantitative limit of pulse durations below which the BSD mechanism will fail? Finally, is there any transition region of pulse durations where possibly both mechanisms coexist or alternatively, do other still unknown mechanisms dominate the dissociation process in this region?

In this work, we will systematically study the dissociation mechanisms of an exactly solvable $\textrm H_2^{\;+}$-like molecular model system for a wide range of pulse durations, extending from $\tau=250$ as up to $\tau=32$ fs. In Sec.~\ref{sec:model}, we will derive explicit expressions for the dissociation probability in the three cases:  first, for the exact solution of the time-dependent Schr\"odinger equation (TDSE), second for the pure VED mechanism and third, for a pure BSD model. In Sec.~\ref{sec:fc_pulses}, we will introduce few-cycle laser pulses and elaborate on their properties. After these preliminary considerations, we are able to clarify the dissociation mechanisms in detail (Sec.~\ref{sec:pdiss_ker_hcp} and \ref{sec:pdiss_ker_fcp}). In Sec.~\ref{sec:attosecond}, we will analyze the dissociation dynamics induced by a realistic, experimentally realized attosecond laser pulse \cite{hassan_2016}. The findings are concluded in Sec.~\ref{sec:conclusion}.

Atomic units (a.u.) are used unless stated otherwise.

\FloatBarrier
\section{Molecular model and dissociation probabilities}\label{sec:model}

\subsection{Molecular model}
All numerical examples presented throughout this work are calculated using a two-dimensional $\textrm H_2^{\;+}$-like molecular model interacting with a linear polarized laser in dipole approximation. This model has a fairly realistic dissociation behavior, and the TDSE for the full electron-nuclear wave function $\psi(R,r,t)$,

\begin{align}\label{eq:TDSE}
\textrm{i}\partial_t\psi(R,r,t)=H(R,r,t)\psi(R,r,t),
\end{align}

can be solved numerically exactly with relatively little effort. In the center-of-mass frame, the Hamiltonian reads 

\begin{align}\label{eq:H_model}
H=&-\frac{\Delta_R}{2M}-\frac{\Delta_r}{2}+\frac{1}{R}-\mu \mathcal{E}(t)\nonumber\\
&-\frac{1}{\sqrt{(r-R/2)^2+1}}-\frac{1}{\sqrt{(r+R/2)^2+1}},
\end{align} 

with a reduced nuclear mass of $M=918$ a.u. $R$ denotes the internuclear distance, and $r$ the electronic position operator. The interaction of the homonuclear molecule (electronic dipole operator $\mu=-r$) with the laser field $\mathcal{E}(t)$ is included in length gauge. The molecular ground state is used as initial state for all calculations. The propagation starts at $t_0=-\tau/2$, such that the laser with pulse duration $\tau$ reaches its peak intensity at $t=0$. The propagation ends at a final time $t_f$ chosen large enough for converged dissociation probabilities.    

To obtain the exact molecular dynamics, the TDSE \eqref{eq:TDSE} is solved using the second order split-operator method with a time step of d$t=0.05$ a.u. (d$t\approx 1.25$ as) on a two-dimensional grid. Combining the time-dependent surface flux method (see \cite{yue_2013} and below) with an absorber to prevent unphysical reflections of the wave function at the grid boundaries allows for the use of a relatively small and efficient $R\times r$ grid with $512\times 256$ grid points spanning from $R=0.05$~a.u. to $R=25$~a.u. for the nuclear bond length and from $r=-30$~a.u. to $r=30$~a.u. for the electronic coordinate. The grid-based propagation of the TDSE allows, in principle, for a detailed investigation of ionization. However, since this work is focused on dissociation, we use only moderate laser intensities and ensured that ionization plays a negligible role. 

For the analysis of the dissociation dynamics during this work, we define BO states $\varphi_i$ and BO Energies $E_i$ as eigenstates and eigenvalues of the field-free BO Hamiltonian $H^{BO}$ (Eq.~\eqref{eq:H_model} for fixed nuclei and $\mathcal{E}=0$), 

\begin{align}
&H^{BO}(R,r)\varphi_i(R,r)=E_i^{BO}(R)\varphi_i(R,r), \\
& \varphi_i(R,r,t)=\varphi_i(R,r)e^{-\textrm{i} E_i^{BO} (R) [t-t_0] },
\end{align}

vibrational states $\chi_{0\nu}$ and vibrational energies $E_{0\nu}$ as (bound) eigenstates and eigenvalues of the ground state BO PES, 

\begin{align}
\left(-\frac{\partial_R^2}{2M}+E_0^{BO}(R) \right)\chi_{0\nu}(R)=E_{0\nu}\chi_{0\nu}(R),
\end{align}

and scattering states $\chi_{1l}$ as well as the corresponding energies $E_{1l}$ as eigenstates and eigenvalues of the first excited BO surface, 

\begin{align}
\left(-\frac{\partial_R^2}{2M}+E_1^{BO}(R) \right)\chi_{1l}(R)=E_{1l}\chi_{1l}(R).
\end{align}

\subsection{Exact dissociation probability from the TDSE}

The (practically) exact dissociation probability density $p_{diss}^{exact}$ as function of the KER is calculated using the time-dependent surface flux method \cite{yue_2013},

\begin{align}\label{eq:tsurff}
p_{diss}^{exact}&\left(\textrm{KER}=\frac{K^2}{2M}\right)=\sum_i \left\{ \frac{2}{\pi M K} \right.   \nonumber \\
& \, \times \left| \int_{t_0}^{t_f}\! \textrm{d}t ~ \exp\left(\textrm{i} \frac{K^2}{2M}[t-t_0] -\textrm{i} K R \right) \right.  \\
&  \left. \left. \times \left[K-\textrm{i}\partial_R \right] \int_{-r_s}^{r_s} \! \textrm{d}r ~ \varphi_i^*(R,r,t)\psi(R,r,t) \right|^2_{R=R_s} \right\}. \nonumber 
\end{align}

The use of the field-free BO states $\varphi_i$ is justified as long as the position $R_s$ of the integration surface is chosen large enough such that the dissociating density reaches $R_s$ at times $t>\tau/2$, when the laser pulse is already over. For a detailed derivation of \eqref{eq:tsurff} and a discussion of $R_s$ and the electronic integration range $-r_s\leq r \leq r_s$  see the original work \cite{yue_2013}.  The dissociation probability $P_{diss}^{exact}$ is  calculated by integration over the KER distribution, 

\begin{align}
P_{diss}^{exact}=\int_0^{\infty} \! \textrm{d} \textrm{KER} ~p_{diss}^{exact}(\textrm{KER}).
\end{align}

For all calculations presented in this work, the first excited BO state, $\varphi_{i=1}$, strongly dominates the dissociation. Contributions of states with $i\neq 1$ are thus neglected for the following discussions. There are no field-free nonadiabatic transitions between $\varphi_0$ and $\varphi_1$, since the corresponding coupling $\int \! \textrm{d}r~ \varphi_1^*(R,r) \partial_R \varphi_0(R,r)$ vanishes exactly. 

For the analysis of the dissociation probability in Sec.~\ref{sec:pdiss_ker_fcp}, we introduce the time-dependent occupation probability of the first excited BO state $P_{occ}^{exact}(t)$,

\begin{align}
P_{occ}^{exact}(t)=\left| \int_0^{\infty} \! \textrm{d}R \int_{-\infty}^{\infty} \! \textrm{d}r ~ \varphi_1^*(R,r) \psi(R,r,t) \right|^2,
\end{align}     

and the time-dependent probability for nuclear excitation in the electronic groundstate $P_{vib}^{exact}(t)$, 

\begin{align}
P_{vib}^{exact}(t)=&\sum_{\nu>0} \left| \int_0^{\infty} \! \textrm{d}R \int_{-\infty}^{\infty} \! \textrm{d}r ~ \right. \nonumber \\
& \left. \phantom{\int_0^{\infty} \! \textrm{d}R} \varphi_0^*(R,r) \chi_{0\nu}^*(R) \psi(R,r,t) \right|^2.
\end{align}  

\subsection{VED probability}\label{sec:pdiss_ved}

The VED mechanism allows for a relatively simple approximative calculation of the  dissociation probability density $p_{diss}^{exact}(KER)$. The vertical excitation of the initial nuclear wave packet $\psi(R,t_0)=\chi_{00}(R)$ yields an excited nuclear wave packet $\Phi(R,t)$, 

\begin{align}\label{eq:vd_expansion}
\Phi(R,t)=\sum_l c_l(t) \chi_{1l}(R) e^{-\textrm{i} E_{1l} [t-t_0]},  
\end{align}       

as superposition of scattering states $\chi_{1l}(R)$ \footnote{We note that the discrete sum is appropriate even for in principle continuous scattering energies, since all our calculations are performed on finite grids anyways.}. Closely following Chap.~$6$ and $13$ in \cite{schinke_1993}, the expansion coefficients are approximated in first order perturbation theory, 

\begin{align}\label{eq:vd_coeffs}
c_l(t)=&-\textrm{i} \int_{t_0}^t \! \textrm{d}t'~ \mathcal{E}(t') e^{\textrm{i}[E_{00}-E_{1l}] t'} \\   
& \times \int_0^{\infty}\! \textrm{d}R~ \mu (R) \chi_{1l}^*(R) \psi(R,t_0). \nonumber
\end{align}

The coefficients are calculated using the explicit shape of the electric field, $\mathcal{E}(t)$, but with the assumption of an undisturbed nuclear wave packet $\psi(R,t)=\psi(R,t_0)$. The dipole moment is $\mu(R)=\int \! \textrm{d}r~ \varphi_1^*(R,r,t_0) \mu \varphi_0(R,r,t_0)$.

Due to the oscillatory nature of the scattering states $\chi_{1l}$, the integral on the second line of \eqref{eq:vd_coeffs} can be approximated by $\mu(R_l)\psi(R_l,t_0)$,  where $R_l$ is the classical turning point $E_1^{BO}(R_l)=E_{1l}$ (see discussion to Fig.~6.2 in \cite{schinke_1993}). This allows for a simple calculation of the occupation probability $P_{occ}^{VED}$ of the first excited BO state as function of time,

\begin{align}\label{eq:occ_ved}
P_{occ}^{VED}(t)&=\sum_l |c_l(t)|^2\\
&=\sum_l \left| \mu(R_l)\psi(R_l,t_0) \int_{t_0}^{t} \! \textrm{d}t'~ \mathcal{E}(t') e^{\textrm{i} [E_{00}-E_{1l}] t'} \right|^2, \nonumber
\end{align}

and the dissociation probability density $p_{diss}^{VED}$,

\begin{align}\label{eq:pdiss_ved}
p_{diss}^{VED} (\textrm{KER}=E_{1l})=|c_l(t_f)|^2,
\end{align}   

as well as the total dissociation probability $P_{diss}^{VED}$,

\begin{align}\label{ved_prob}
P_{diss}^{VED}=\sum_l |c_l(t_f)|^2.
\end{align}

\subsection{BSD probability}

To obtain the dissociation probability for the BSD mechanism, the initial nuclear wave packet $\psi(R,t_0)=\chi_{00}(R)$ is propagated on the bond softening Floquet surface (see red line in Fig.~\ref{fig:fcp_bs}). This Floquet surface is time-dependent via the envelope of the various laser pulses (see, e.g., \cite{fiedlschuster_2016}). At the end of the laser pulse, $t=\tau/2$, the dissociation probability $P_{diss}^{BSD}$ is calculated as 

\begin{align}\label{eq:pdiss_bs}
P_{diss}^{BSD}=1-\int_{0}^{R_{b}} \! \textrm{d}R~  |\psi(R,\tau/2)|^2,
\end{align}

where $R_b$ is the position of the dissociation barrier of the Floquet surface for vanishing laser intensity ($R_{b}\approx 3.3$ a.u. for the employed $\omega_l=0.2$ a.u.). 

The restriction to a single Floquet surface is an idealization suitable to understand and discuss the BSD mechanism. Replacing the exact solution of the two-dimensional TDSE by wave packet dynamics on Floquet surfaces requires, in principle, coupled dynamics including (infinitely) many Floquet surfaces, with an exception only given by the case of an exact continuous wave laser and special initial conditions. In the presented case, the restriction to a single surface manifests in a slight dissociation even in field-free benchmark calculations, since the initial state is not a stationary state of the bond softening Floquet surface. This unphysical behavior can be substantially corrected by including one additional Floquet surface, the surface relevant for bond hardening or vibrational trapping (see, e.g., Fig.~2 in \cite{ibrahim_2017}). For simplicity, however, we stick to the propagation on the single surface and slightly correct the dissociation probability \eqref{eq:pdiss_bs} by subtracting the unphysical field-free dissociation probability, which takes values between $0.3\%$ and $7\%$, depending on the pulse duration $\tau$.

\section{Few-cycle laser pulses}\label{sec:fc_pulses}

Before applying any model calculations, it is instructive to recognize the general features of the employed very short laser pulses. The exact definition of the electric field $\mathcal{E}(t)$, its properties and, in particular, its intensity spectral density $|\mathcal{E}(\omega)|^2$ (simply called frequency spectrum or just spectrum in the following),

\begin{align}\label{eq:spektrum}
|\mathcal{E}(\omega)|^2&=\frac{1}{2\pi} \left| \int \! \textrm{d}t ~ \mathcal{E}(t) e^{-\textrm{i} \omega t} \right|^2, 
\end{align}

are useful information to understand already qualitatively some of the expected phenomena. The spectrum is a direct measure of the strength with which a certain frequency $\omega$ is represented in the laser pulse. It is the central quantity to characterize short laser pulses \footnote{As soon as multi-photon effects play a role, the full (complex) Fourier transform $\mathcal{E}(\omega)$, or a combination of amplitude spectrum and phase spectrum, has to be considered. This is not the case in this work.} and gives first hints to the expected, respectively to the allowed, dissociation mechanism. This will be detailed in the following.

For conventional femtosecond pulses with carrier frequencies in or near the range of the visible spectrum (a frequency of $\omega=0.1$ a.u. corresponds to a wave length of $\lambda\approx 450$ nm), the electric field $\mathcal{E}$ is usually treated in dipole approximation and written as a product of an  envelope function $F(t)$ and a plane carrier wave with a carrier frequency $\omega_l$ and a carrier envelope phase (CEP) $\varphi$,
 
\begin{align}\label{eq:fcp_cpulse}
\mathcal{E}(t)=F(t) \cos \left( \omega_l t +\varphi \right).
\end{align}

The envelope $F(t)$, which determines the pulse duration $\tau$, contains a large number of optical cycles, and changes only marginally on the time scale of one optical cycle $T=2\pi/\omega_l$. For such pulses, the CEP $\varphi$ does not play a significant role and the spectrum $|\mathcal{E}(\omega)|^2$ is sharply peaked at the laser frequency $\omega=\omega_l$. These attributes are ideal premises to apply Floquet theory in the description of the molecule-laser interaction.

When advancing into the regime of very short femtosecond pulses, the use of \eqref{eq:fcp_cpulse}, however, results in a considerable conflict with Maxwell's equations for a freely propagating electromagnetic field of finite duration (see, e.g., \cite{madsen_2002}). This conflict is accompanied by a non-negligible unphysical direct current component in the electric field of the laser, which will adulterate the molecular dynamics. Maxwell's equations require equal values of the vector potential $A(t)$ at the beginning and the end of the laser pulse, or equivalently, a vanishing time integral over the electric field $\mathcal{E}(t)$, usually called zero net force condition \cite{madsen_2002}. Thus, for the rest of this work, we define the electric field of any pulse as
 
\begin{align}\label{eq:fcp_efield}
\mathcal{E}(t)=-\partial_t A(t),
\end{align}  

with the vector potential 

\begin{align}\label{eq:fcp_vecpot}
A(t)=A_0 \cos^2 \left(\frac{\pi t}{\tau} \right) \cos \left( \omega_l t +\varphi+\frac{\pi}{2} \right)
\end{align} 

for $-\tau/2\leq t \leq \tau/2$, and $A(t)=0$ otherwise. This ensures that the resulting pulses satisfy the zero net force condition exactly. The CEP is restricted to $\varphi=0$ for the following investigations \footnote{We have checked that the main conclusions drawn in this work also hold for other values of the CEP.}.

The total pulse duration $\tau$ is conveniently expressed using the (not necessarily integer) number of optical cycles $n$ contained in the pulse envelope, 

\begin{align}
\tau=n\times T.
\end{align}

It is easily shown that for large $\tau$, respectively large $n$, the electric field $\mathcal{E}(t)$ resulting from \eqref{eq:fcp_vecpot} approaches the approximate expression \eqref{eq:fcp_cpulse}.  It is also easy, at least theoretically, to systematically regulate the total pulse duration by varying the period $T$ via the carrier frequency $\omega_l$ for fixed $n$, or by changing the number of optical cycles $n$ for a fixed period $T$. Both cases are presented in Figs.~\ref{fig:hcp_laser} and~\ref{fig:fcp_laser} and will be discussed in the following.

\subsection{Half-cycle pulses ($\mathbf{n=1/2}$ and $\mathbf{\omega_l=0.05,\ldots, 0.3}$ a.u.)}\label{sec:discussion_hcp}

\begin{figure}[h]
\begin{center}
\includegraphics[scale=0.5]{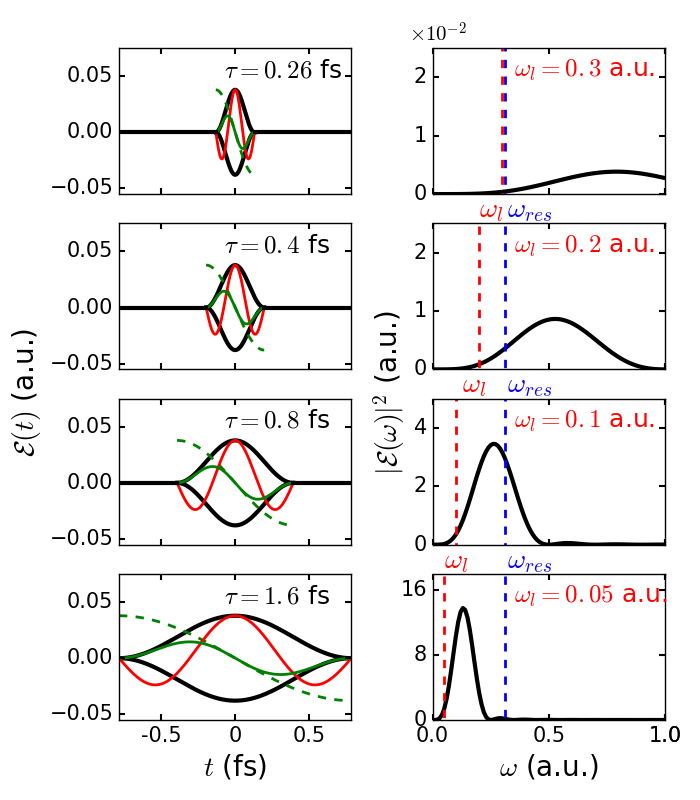}
\caption{Left panels: The electric field (red line), the vector potential (green line), and the $\cos^2$-shaped envelope of the vector potential (black line) for half-cycle pulses defined with different carrier frequencies $\omega_l=0.3$ a.u., $0.2$ a.u., $0.1$ a.u., and $0.05$ a.u. The peak intensity is $I=5\times10^{13}$ W/cm$^2$ for all pulses.  The vector potential and its envelope are scaled by a factor of $\omega_l$. The dashed green line shows the vector potential without modulation by the envelope. Right panels: The frequency spectrum of the pulses shown in the left panels.  The dashed vertical lines mark the carrier frequencies $\omega_l$ of the laser (red) and the mean value of the VED resonance frequencies $\omega_{res}=0.31$ a.u. of the molecule (blue).} 
\label{fig:hcp_laser}
\end{center}
\end{figure}

In Fig.~\ref{fig:hcp_laser}, the electric field \eqref{eq:fcp_efield}, the vector potential \eqref{eq:fcp_vecpot}, and the frequency spectrum \eqref{eq:spektrum} for half-cycle pulses defined with different carrier frequencies $\omega_l$ is shown. The laser intensity is $I=(\omega_l A_0)^2/2=5\times 10^{13}$ W/cm$^2$ for all pulses. 

As seen in the left panels of Fig.~\ref{fig:hcp_laser}, the envelope of the vector potential heavily affects the shape of the electric field for two reasons. First, the envelope qualitatively changes the carrier wave of the vector potential (c.f. dashed green lines and full green lines in Fig.~\ref{fig:hcp_laser}). Second, the time derivative of the envelope even dominates over the time-derivative of the carrier wave when calculating the electric field via \eqref{eq:fcp_efield}.

Not surprisingly, the short half-cycle pulses do have an extremely broad frequency spectrum. With increasing pulse duration, the spectrum shows a clear tendency to gradually form a distinct maximum at smaller and smaller frequencies, shifting towards $\omega_l$ (right panels from top to down in Fig.~\ref{fig:hcp_laser}).  Nevertheless, the dominant contributions of $|\mathcal{E}(\omega)|^2$  do not have remarkable overlap with $\omega_l$. In contrast, the overlap between the spectrum and the resonance frequencies $\omega_{res}$ of VED depends strongly on the pulse duration and reaches a maximum at $\tau=0.8$ fs, respectively $\omega_l=0.1$ a.u. Hence, one can expect a maximum in the VED probabilities at these values. Note that the spectrum directly enters the VED probability \eqref{ved_prob}.

Summarizing this part, for half cycle pulses, the variation of the pulse duration $\tau=1/2\times T$ via the period of an optical cycle $T= 2\pi/\omega_l$ does not change the general property of the pulses as a broad distribution over many frequencies. The variation of $\omega_l$, however, changes the pronunciation of the frequencies relevant for the VED mechanism,  $\omega\approx \omega_{res}$, in the frequency spectrum $|\mathcal{E}(\omega)|^2$. One may expect distinct structures in the final VED probabilities as function of the pulse duration $\tau$, respectively the carrier frequency  $\omega_l$, and VED generally as the dominating or only dissociation mechanism. The carrier frequency $\omega_l$ of the laser is merely a parameter to define the pulse duration, rather than a physical relevant quantity. It is not directly recognized in the resulting vector potential and certainly not in the electric field.

Finally, we note that half-cycle pulses, defined with $n=1/2$ in the vector potential \eqref{eq:fcp_vecpot}, indeed do have some selected properties. Among other things, the qualitative form of the electric field as shown in Fig.~\ref{fig:hcp_laser} remains unchanged if even shorter pulses with $n=1/4$, $1/8$, $\ldots$ (not shown) are considered. In addition, this form of the electric field can serve as a good theoretical approximation and idealization to realistic, experimentally achievable attoseconds pulses (see Sec.~\ref{sec:attosecond}).

\subsection{Pulses with increasing number of optical cycles ($\mathbf{n=1/2,1,\ldots,10}$ and $\mathbf{\omega_l=0.2}$ a.u.)}\label{sec:discussion_fcp}

\begin{figure}[h]
\begin{center}
\includegraphics[scale=0.5]{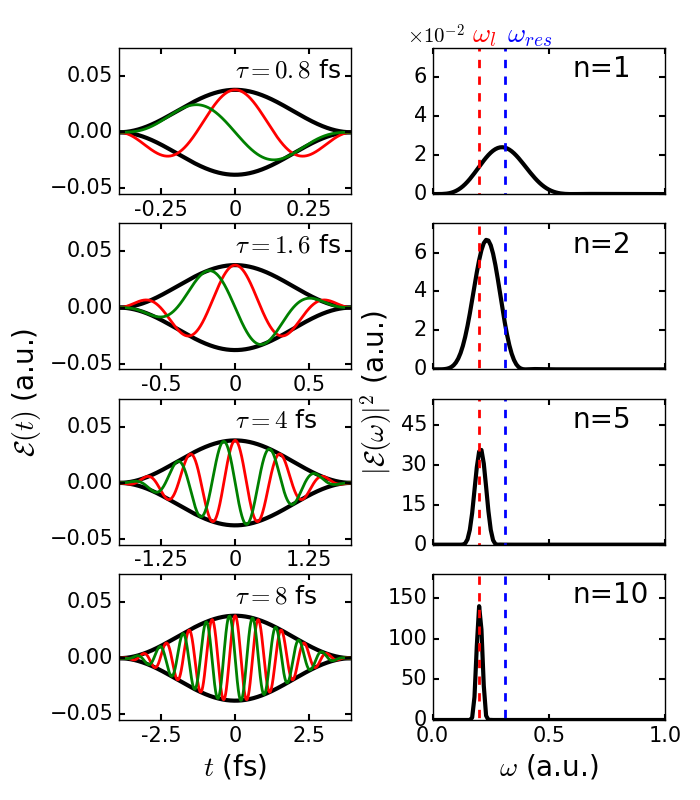}
\caption{The same as in Fig.~\ref{fig:hcp_laser}, but varying the pulse duration $\tau$ by the number of optical cycles $n$ at a fixed carrier frequency of $\omega_l=0.2$ a.u., respectively an optical period of $T=0.8$ fs. Not shown are the dashed green lines.} 
\label{fig:fcp_laser}
\end{center}
\end{figure}

In Fig.~\ref{fig:fcp_laser}, the same as in Fig.~\ref{fig:hcp_laser} is shown but for few-cycle pulses with increasing number of optical cycles ($n=1$, $2$, $5$, and $10$), and at a fixed carrier frequency of $\omega_l=0.2$ a.u., respectively an optical period of $T= 0.8$ fs. Also the half-cycle pulse is included in the following discussion (the second row of Fig.~\ref{fig:hcp_laser} complements the graphic of Fig.~\ref{fig:fcp_laser}  with $n=1/2$).

As clearly seen when comparing the half-cycle pulse (Fig.~\ref{fig:hcp_laser}) and the one-cycle pulse (Fig.~\ref{fig:fcp_laser}), the step from $n=1/2$ to $n=1$ already significantly reduces the effect of the envelope on the carrier wave. With further increasing pulse duration (respectively $n=2$, $5$, $10$), the electric field gets more and more reminiscent of a continuous wave laser, as the envelope, both its change during one optical cycle and its time derivative, becomes less significant (see left panels in Fig.~\ref{fig:fcp_laser}). For pulses with $n\gtrsim 10$, the time-derivative of the envelope barely contributes to the electric field, thus \eqref{eq:fcp_efield} and \eqref{eq:fcp_vecpot} are nearly identical to the electric field \eqref{eq:fcp_cpulse} introduced for conventional femtosecond pulses at the beginning of this section.

Accordingly, the broad frequency spectrum of the half-cycle pulse quickly evolves into a distinctly peaked distribution already for $n=1$, and is sharply peaked around $\omega_l$ for $n=5$ and $n=10$. For even larger $n$, the spectrum finally reaches a delta-function-like form, $|\mathcal{E}(\omega)|^2\sim\delta(\omega-\omega_l)$, which is physically the ultimate limit and characterizes a continuous wave laser. At the same time, it signals the eligibility to apply Floquet theory. On the way to this limit, the frequency spectrum passes through the region of $\omega_{res}$ with different strengths for $n=1/2$, $1$, $2$, and $n=3$, $4$ (not shown) with a pronounced maximum overlap for $n=1$ and $n=2$. Hence, one may expect a maximum in the VED probability at $n=1$ and $n=2$, reflecting the large contribution of frequencies around $\omega_{res}$ to the spectrum of these pulses.

Summarizing this section, the variation of $\tau=n\times T$ by increasing the number of optical cycles $n$, drastically changes the properties of the pulses. For small values of $n$, i.e. $n \gtrsim 1$, the pulses still have the same general features as half-cycle pulses. Hence, the same dissociation mechanism can be expected, including distinct structures in the final VED probabilities as function of $\tau$. However, above a certain value of $n$, the spectrum of the pulses sharply peaks exactly at the carrier frequency $\omega_l$, unambiguously indicating that Floquet theory can be applied. To what extend this change in the spectrum really leads to a change in the dissociation mechanism, eventually from VED to BSD, depends on the time scale of the nuclear motion which must be comparable with the actual short  pulse durations. This, however, can be probed only by dynamical model calculations of the full molecule-laser interaction which will be presented, inter alia, in the next section.

\section{Dissociation in few-cycle pulses}\label{sec:results}

In this central part of the paper, the dissociation probability and the probability density (i.e. the KER spectra of the fragments) of the $\textrm H_2^{\;+}$-like molecular model exposed to few-cycle laser pulses are discussed. First, the exact solution of the TDSE will be presented, delivering indeed surprising results, which can hardly be explained by the TDSE alone. Second, to hopefully obtain a physical transparent picture of the dissociation pattern, the exact results are then analyzed by model calculations of the VED and BSD mechanisms.
 
The same classification of the pulses as introduced in the previous section is used, i.e. half-cycle pulses (Sec.~\ref{sec:discussion_hcp}) and pulses with increasing number of optical cycles (Sec.~\ref{sec:discussion_fcp}).  The pulse durations range from several attoseconds up to a few femtoseconds. All examples of pulses,  characterized in Figs. \ref{fig:hcp_laser} and \ref{fig:fcp_laser}, are included.

\subsection{Half-cycle pulses}\label{sec:pdiss_ker_hcp}

\begin{figure}[htb]
\begin{center}
\includegraphics[scale=0.5]{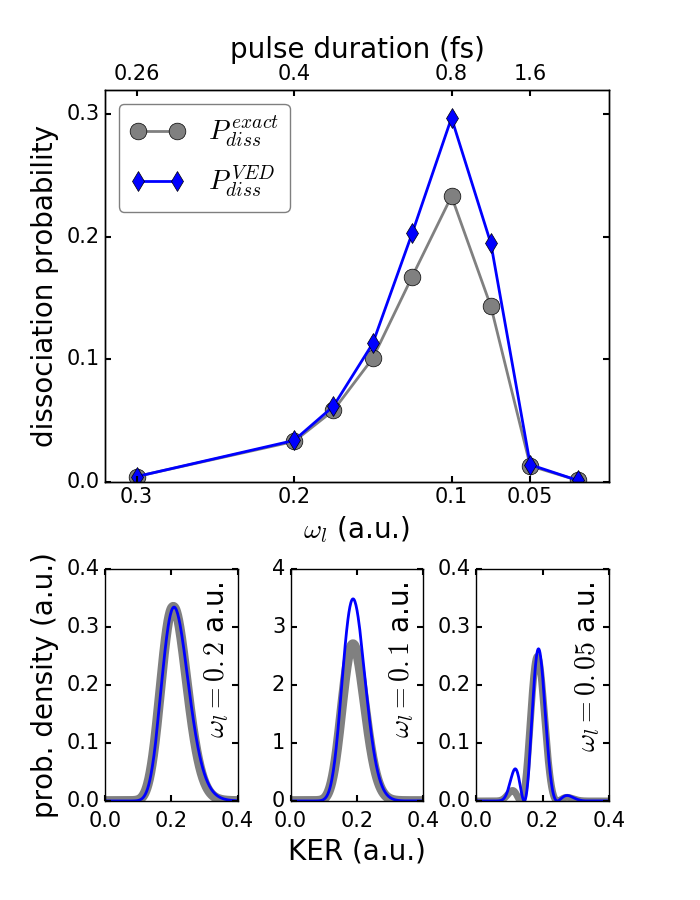}
\caption{Dissociation in half-cycle pulses. Upper panel: the exact dissociation probability (gray circles) and the VED probability (blue diamonds) for half-cycle pulses as function of the carrier frequency $\omega_l$ (lower scale), respectively the pulse duration $\tau$ (upper scale). The laser intensity of $I=5\times 10^{13}$ W/cm$^2$ is the same as used in Sec.~\ref{sec:fc_pulses}. Lower panels: the exact dissociation probability density (gray lines) and the VED probability density (blue lines) for selected half-cycle pulses with different $\omega_l$ (respectively $\tau$), i.e. $\omega_l=0.2$ a.u.($\tau=0.4$ fs), $\omega_l=0.1$ a.u ($\tau=0.8$ fs) and $\omega_l=0.05$ a.u. ($\tau= 1.6$ fs).} 
\label{fig:hcp_pdiss}
\end{center}
\end{figure}

In Fig.~\ref{fig:hcp_pdiss}, the dissociation probability and probability density for different half-cycle pulses, defined with the same carrier frequencies and the same intensity of $I=5\times 10^{13}$ W/cm$^2$ as used in Sec.~\ref{sec:fc_pulses}, is shown.

The numerically exact dissociation probability strongly depends on $\omega_l$, with a pronounced peak at $\omega_l=0.1$ a.u. (see upper panel of Fig.~\ref{fig:hcp_pdiss}). From the point of view of the TDSE, such a behavior is completely surprising and a physical explanation of this phenomenon is lacking.

On the other side, this is exactly the behavior expected from the evolution of the spectrum of the half-cycle pulses as function of $\omega_l$ (respectively $\tau$), presented in Sec.~\ref{sec:discussion_hcp}.  For $\omega_l=0.1$ a.u. ($\tau=0.8$ fs), the spectrum peaks at the resonance frequencies for VED of the molecule, $\omega\approx\omega_{res}$, while other pulses yield smaller contributions at the frequencies around $\omega_{res}$ (see Fig.~\ref{fig:hcp_laser}). This necessarily leads to larger VED excitation and dissociation probabilities for  $\omega_l \approx 0.1$ a.u. (i.e. $\tau \approx 0.8$ fs) than for all other carrier frequencies (respectively pulse durations). Thus, VED is  strongly suggested as the responsible dissociation mechanism.

Indeed, the approximate VED probability (blue diamonds in the upper panel of Fig.~\ref{fig:hcp_pdiss}) is qualitatively in very good agreement with the exact result. The quantitative overestimation of the VED probability around $\omega_l=0.1$ a.u. is not surprising but rather expected, since this  probability is calculated in first order perturbation theory, where the basic assumption of an undisturbed ground state wave packet (see Eq.~\eqref{eq:vd_coeffs}) becomes less accurate for the noticeable dissociation probabilities observed at these frequencies.

In the lower panels of Fig.~\ref{fig:hcp_pdiss}, the dissociation probability densities for three exemplary half-cycle pulses are shown. In all cases, the KER spectrum consist of a dominating peak centered at KER $\approx \omega_{res}-E_b=0.2$ a.u. This is predicted by the exact TDSE results, qualitatively expected from the VED mechanism (see Fig.~\ref{fig:fcp_bs}), and described quantitatively also by the VED calculations. In addition, the VED model describes even teeny details of the exact TDSE result. This concerns e.g. the shape and the exact position of the main peak, which moves slightly, but still observable, to smaller KER energies with decreasing $\omega_l$. In addition, for the "longest" pulse with $\omega_l=0.05$ a.u., small additional peaks appear on each side of the main peak. All these trifles can be understood and interpreted by exploring the detailed properties of the spectra of the pulses together with their responsible influence to generate the VED model probabilities. Here, however, we will pass on these elaborate considerations.

In summary, the dissociation of the idealized $\textrm H_2^{\;+}$-like molecular model exposed to half-cycle pulses with parameters in and near the optical range generally and exclusively proceeds via the VED mechanism. A simple VED approximation reproduces in detail the dissociation probability and the KER spectra predicted by the TDSE. First and foremost, VED allows for a transparent explanation of the manifold dissociation phenomena.

\FloatBarrier
\subsection{Pulses with increasing number of optical cycles}\label{sec:pdiss_ker_fcp}

\begin{figure}[htb]
\begin{center}
\includegraphics[scale=0.5]{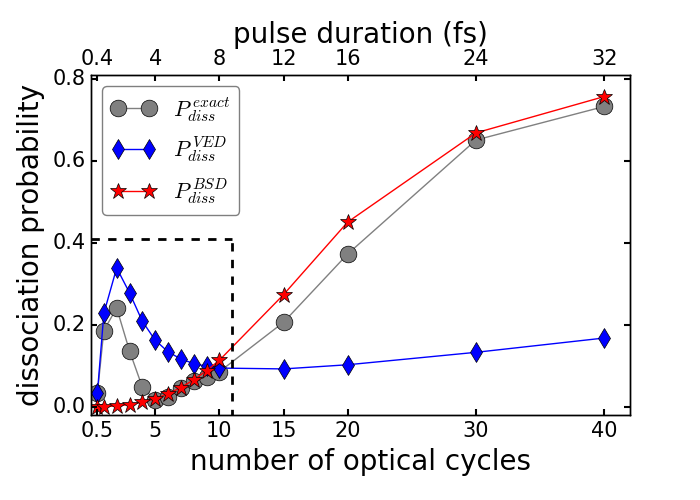}
\caption{Dissociation in pulses with increasing number of optical cycles. The dissociation probabilities are calculated from the TDSE (gray circles), the VED model (blue diamonds), and the simplest BSD model (red stars). The laser frequency is fixed to $\omega_l=0.2$ a.u. ($T=0.8$ fs), the intensity is $I=5\times 10^{13}$ W/cm$^2$. The number of optical cycles ranges from $n=1/2$ to $n=40$ (lower scale), which corresponds to pulse durations from $\tau=0.4$ fs to $\tau=32$ fs (upper scale). The region with $n\leq 10$ (area marked by the dashed lines) is shown on a larger scale in Fig.~\ref{fig:pdiss_n2}.} 
\label{fig:pdiss_n1}
\end{center}
\end{figure}

\begin{figure}[htb]
\begin{center}
\includegraphics[scale=0.5]{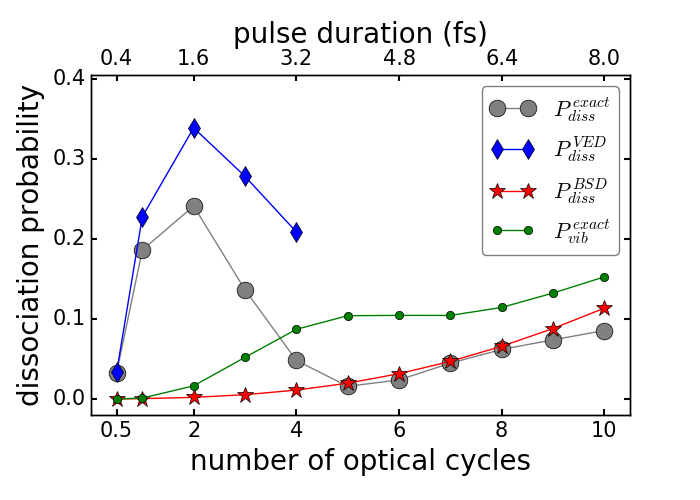}
\caption{The various dissociation probabilities in the transition region between VED and BSD shown on a larger scale as presented in Fig.~\ref{fig:pdiss_n1}.  The additionally given occupation probability of excited vibrational states at the end of the pulse $P_{vib}^{exact}(t_f)$ (green dots) signals the failure of the fixed-nuclei approximation during the laser (see text).}
\label{fig:pdiss_n2}
\end{center}
\end{figure}

The various dissociation probabilities, calculated for few-cycle pulses defined with a fixed carrier frequency of $\omega_l=0.2$ a.u., a peak intensity of  $I=5\times 10^{13}$ W/cm$^2$, but different pulse durations $\tau$ characterized by the number of optical cycles $n\leq 40$, are presented in Figs.~\ref{fig:pdiss_n1} and \ref{fig:pdiss_n2}. For longer pulses, $n>40$, the rising ionization probability starts to suppress the dissociation. Thus, Fig.~\ref{fig:pdiss_n1} contains the whole relevant range of pulse durations for pure dissociation of the employed model system.

The exact dissociation probability shows a significant increase from half-cycle pulses with $n=1/2$ to $n=1$ and $n=2$. It drops down for $n=3$ and $n=4$, until it reaches a pronounced minimum for the five-cycle pulse $n=5$. After that, the dissociation probability steadily rises with increasing $n$.

Again, but yet somewhat noteworthy, the gross features of this behavior are already expected from the evolution of the spectrum with respect to the distinct frequencies $\omega_l$ and $\omega_{res}$ (see Fig.~\ref{fig:fcp_laser}). Instead of repeating the qualitative arguments given in Sec.~\ref{sec:fc_pulses}, we will give a detailed discussion of the applicability of the two basic mechanisms, VED and BSD, to interpret the exact dissociation probability.

Starting with VED, $P_{diss}^{VED}$ shows a good agreement with the exact result for the shortest pulses $n=1/2$ and $n=1$ (see Fig.~\ref{fig:pdiss_n2}).  It also reproduces the peak around $n=2$, respectively the drop for $n\geq 3$, at least qualitatively.  The quantitative overestimation of $P_{diss}^{VED}$ for $n \geq 3$, however, is not a simple accuracy problem arising from first order perturbation theory. Instead, it is a direct consequence of leaving the scope of the VED mechanism, which is strongly restricted to a practically instantaneous electronic excitation during the laser pulse (see Sec.~\ref{sec:intro} and \ref{sec:pdiss_ved}). This condition is considerably violated already for $n\geq3$, as indicated by the non-negligible excitation probability of vibrational motion at the end of the pulse (see green line in Fig.~\ref{fig:pdiss_n2}), and as will be discussed further towards the end of this section.

The BSD mechanism, on the other side, describes the exact dissociation probability nearly perfectly, starting at $n = 5$ (see Figs.~\ref{fig:pdiss_n1} and \ref{fig:pdiss_n2}). The perfect agreement is indeed a remarkable result since we used the simplest Floquet-based model, i.e. nuclear motion on a single Floquet surface instead on an infinite number of coupled Floquet surfaces mediated by non-adiabatic transitions \cite{fiedlschuster_2016,fiedlschuster_2017}.

Another astonishing observation concerns the applicability and nearly exact validity of the Floquet-theory based BSD mechanism down to $n=5$, respectively down to a pulse duration of $\tau=4$ fs. This is unexpected, since Floquet theory is, in principle, exact only for continuous wave lasers, respectively for $n\to \infty$ and $\tau \to \infty$, in our context \cite{shirley_1965}.

\begin{figure}
\begin{center}
\includegraphics[scale=0.5]{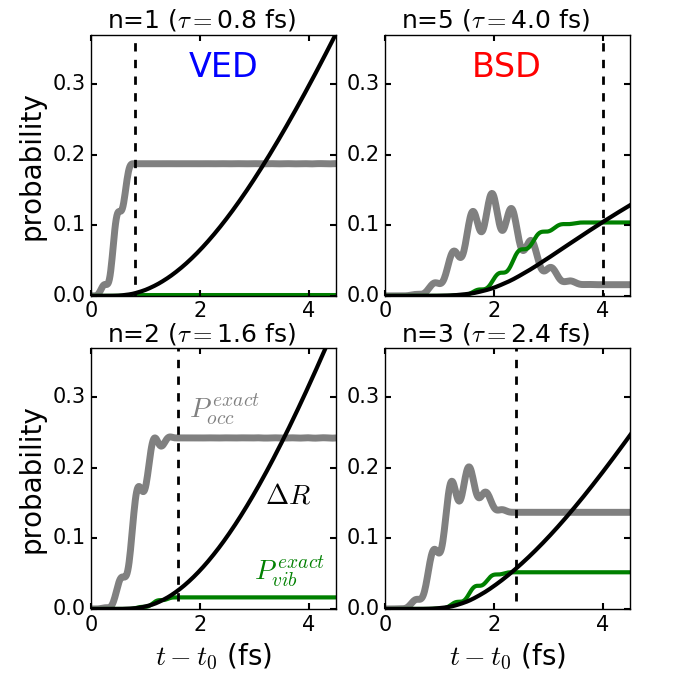}
\caption{Visualization of the coupled electron-nuclear dynamics: The occupation probability of the first excited BO state $P_{occ}^{exact}$ (gray line), the probability for vibrational excitation $P_{vib}^{exact}$  (green line), and the change of the molecular bond length $\Delta R$ (black line, in a.u.) as function of time during selected laser pulses ($n=1,2,3,5$).} 
\label{fig:occ}
\end{center}
\end{figure}

To convey a better understanding of the transition from VED to BSD, the coupled laser-driven electron-nuclear dynamics is in extenso and convincing presented in Fig.~\ref{fig:occ}, where the occupation probability of the first excited BO state $P_{occ}^{exact}$, the vibrational excitation probability $P_{vib}^{exact}$, and, as a coarse measure of nuclear motion, the increase of the expectation value of the molecular bond length $\Delta R= \langle R \rangle (t)-\langle R \rangle (t_0)$, with

\begin{align}
\langle R \rangle (t)= \int_0^{\infty} \! \textrm{d}R \int_{-\infty}^{\infty} \! \textrm{d}r ~ R |\psi(R,r,t)|^2,
\end{align}

are shown as function of time for selected pulses.

For the one-cycle pulse (upper left panel in Fig.~\ref{fig:occ}), the steadily rising $P_{occ}^{exact}$ is well approximated by the assumption of an instantaneous excitation, and nuclear motion during the laser pulse is negligible. Thus, $n=1$ represents a prototype of the VED mechanism. 

On the contrary, for $n=5$ (upper right panel in Fig.~\ref{fig:occ}), a large part of the electronic excitation achieved during the rising edge of the pulse is subsequently transferred into vibrational motion while the laser is still active. These comparably complicated electron-nuclear dynamics are a clear sign that VED is not applicable. On the other hand, the perfect agreement between $P_{diss}^{exact}$ and $P_{diss}^{BSD}$ observed for $n\geq 5$ (see Figs.~\ref{fig:pdiss_n1} and \ref{fig:pdiss_n2}), convincingly suggests BSD as relevant dissociation mechanism \footnote{For lucidity, we intentionally pass on an extensive analysis of the electron-nuclear dynamics in the Floquet picture.}.

In the intermediate region, $1<n<5$, the premises for a clean application of VED are not given, yet the exact dissociation probability is considerably larger than proposed by the simple BSD model. With increasing pulse duration, the dynamics shift from an instantaneous excitation towards an excitation with subsequent conversion into nuclear motion already during the pulse (see lower panels in Fig.~\ref{fig:occ} for $n=2$ and $n=3$). This qualitative change of the electron-nuclear dynamics during the pulse, rather smooth when discussed in terms of the exact TDSE, manifests in the crude failure of both employed dissociation models, VED and BSD, for $1<n<5$ (see Fig.~\ref{fig:pdiss_n2}).

Summarizing the whole Sec.~\ref{sec:results}, the present investigation of the dissociation mechanism of a H$_2^+$-like molecular model revealed three regimes with respect to the pulse duration. First, for very short pulses, $\tau\lesssim 1$ fs, the well-known VED mechanism, where the laser prepares an excited nuclear wave packet which then dissociates after the pulse, yields a qualitatively and quantitatively reliable description of the dissociation dynamics. Second, for relatively long pulses, astonishingly starting at $\tau=4$ fs in the presented examples, the BSD mechanism applies, where a nuclear wave packet dissociates on a single Floquet PES already during the pulse. Third, in the intermediate region $1$ fs $<\tau<4$ fs, a clean discussion in terms of VED or BSD fails, since the premises for both model mechanisms are to heavily violated. The exact solution of the TDSE, however, reveals that a relatively smooth transition between both mechanisms occurs in this intermediate region.  

Especially for the extremely short half-cycle pulses of Sec.~\ref{sec:discussion_hcp}, the minute and explicit consideration of the specific laser pulse was important to understand the dissociation dynamics.  In the next step, we will check if the so far employed laser pulses can serve as valid model for realistic, i.e. experimentally achieved, attosecond laser pulses.

\FloatBarrier
\section{Dissociation in realistic attosecond pulses}\label{sec:attosecond}


During the last years, different methods have been developed to advance the technology of pulsed lasers into the attosecond regime \cite{paul_2001,krausz_2009,leone_2016,hassan_2016,nisoli_2017}. In \cite{hassan_2016}, an optical attosecond pulse was constructed by the superposition of four femtosecond pulses with frequencies ranging from near infrared to ultraviolet wave lengths, thus spanning over (and slightly beyond) the whole visible spectrum. By a clever combination of the single pulses, the authors of \cite{hassan_2016} managed to shorten the (full width at half intensity maximum) duration of the resulting laser pulse to $\tau=380$ as.

In this section, the laser pulse from \cite{hassan_2016} will be compared to an idealized half-cycle pulse as defined in Sec.~\ref{sec:fc_pulses}. A direct comparison of the electric fields and the resulting dissociation dynamics will determine to what extent the model pulse can serve as approximation of the realistic pulse.

\begin{figure}[htb]
\begin{center}
\includegraphics[scale=0.5]{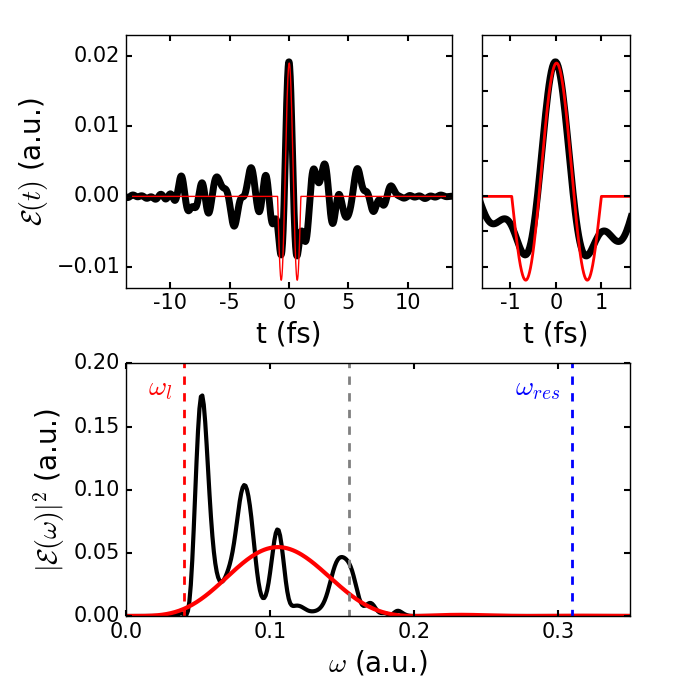}
\caption{Comparison of the electric field (upper panels) and the amplitude spectrum (lower panel) of the experimental attosecond pulse from \cite{hassan_2016} (black line) and the half-cycle pulse calculated from \eqref{eq:fcp_vecpot} (red line). The dashed red line marks the carrier frequency $\omega_l$ used to define the half-cycle pulse. The dashed blue line marks the molecular resonance frequency $\omega_{res}$, the dashed gray line at $\omega=\omega_{res}/2$ marks $\omega_{res}$ for the frequency-doubled pulses (see text).} 
\label{fig:laser2}
\end{center}
\end{figure}

In the upper panels of Fig.~\ref{fig:laser2}, the experimentally constructed and measured electric field of the attosecond pulse is compared to a half-cycle pulse defined by \eqref{eq:fcp_vecpot} with $n=1/2$ and $\omega_l=0.04$ a.u. The half-cycle pulse yields a surprisingly good approximation of the electric field in the central peak of the optical attosecond pulse. A decisive difference is the non-vanishing intensity in the temporal wings of the experimental laser. Albeit much weaker than in the central peak,  the non-zero electric field might still (considerably) affect the molecular dynamics during a time window of nearly $20$ fs. It is thus an interesting question, if the VED mechanism, which covers the dissociation for the half-cycle pulse (see Sec.~\ref{sec:discussion_hcp}), also applies in the case of the experimental attosecond pulse despite the complicated electric field in the pulse wings.  

As a side remark, we note that pulses like the experimental one shown in Fig.~\ref{fig:laser2} are also  frequently called half-cycle pulses, since they are dominated by the electric field reminiscent of half an optical cycle of a certain carrier frequency in the central peak. This (self-evident) characterization as half-cycle pulse is very different from the notation used in this work (see Sec.~\ref{sec:fc_pulses}).

In the lower panel of Fig.~\ref{fig:laser2}, the spectrum of the experimental pulse is compared to the spectrum of the half-cycle pulse. Both spectra cover the same frequency range, but are qualitatively quite different. The diverse spectra clearly adumbrate differing molecular dynamics. 

First of all, however, the spectrum of both pulses lacks noticeable contributions in the vicinity of $\omega_{res}=0.31$ a.u. (see lower panel in Fig.~\ref{fig:laser2}). It is immediately clear that both pulses are not able to dissociate the employed model system via VED. Indeed, the solution of the full TDSE confirms that the molecular system barely responds to either of the pulses, despite the peak intensity of $I=\mathcal{E}^2(t=0)/2=1.25 \times 10^{13}$ W/cm${}^2$ being in principle sufficiently high to dissociate the molecular model.  

To get a significant reaction while maintaining the overall shape of the electric field, we decided to stretch the spectrum of both pulses by a factor of two. The stretched spectrum results in even shorter pulses and also shrinks the temporal wings of the experimental pulse by a factor of two, but now provides contributions around $\omega_{res}=0.31$ a.u. (see dashed gray line in Fig.~\ref{fig:laser2}). Solving the full TDSE for the frequency doubled pulses yields a dissociation probability of $P_{diss}^{exact}\approx 0.05$ in both cases, thus the model \eqref{eq:fcp_vecpot} is a valid approximation for the realistic laser with regard to this integral  observable.

\begin{figure}[htb]
\begin{center}
\includegraphics[scale=0.5]{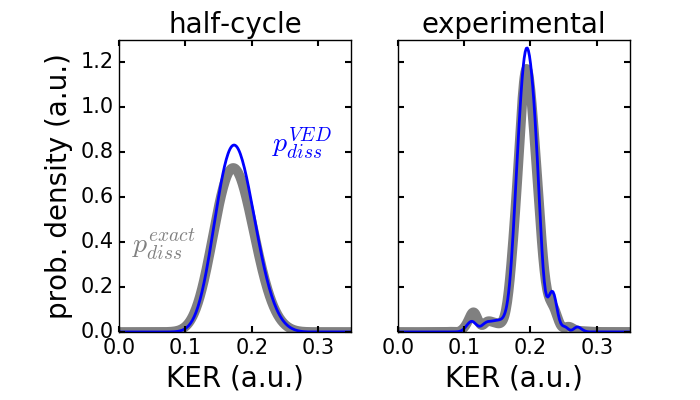}
\caption{The dissociation  probability density for the half-cycle pulse defined with $\omega_l=0.08$ a.u. (left panel) and the frequency-doubled experimental pulse (right panel) calculated from the TDSE (gray line) and with the VED model \eqref{eq:pdiss_ved} (blue line).} 
\label{fig:laser2a}
\end{center}
\end{figure}

The dissociation probability densities $p_{diss}^{exact}(KER)$ resulting from the frequency doubled pulses are shown in Fig.~\ref{fig:laser2a}. The half-cycle pulse yields the rather featureless peak already discussed in Sec.~\ref{sec:pdiss_ker_hcp}. The frequency-doubled experimental pulse results in a comparably narrow main peak, accompanied by some structure in its wings. Once again, this can be discussed with the spectrum in the vicinity of $\omega_{res}$ (see lower panel in Fig.~\ref{fig:laser2}). The monotonous spectrum of the half-cycle pulse results in one featureless peak, while the spectrum of the frequency-doubled experimental pulse peaks at $\omega\approx \omega_{res}$ itself, leading to a narrower main peak in the probability density. The substructure in the wings of the density peak is directly related to the structure of the spectrum around $\omega_{res}$. 

As further seen in Fig.~\ref{fig:laser2a}, the VED probability density is in very good agreement with the exact result, for both pulses. Thus, while the temporal wings of the experimental pulse \emph{do} affect the dissociation dynamics, it is not due to nuclear motion during these temporal wings, but due to their effect on the specific frequency spectrum of the pulse.         

Summarizing this section, the idealized electric field resulting from \eqref{eq:fcp_vecpot} is a useful approximation to the realistic pulse, on a basic level. The model pulse yields the dissociation probability and the gross position and shape of the KER spectrum of the experimental pulse with reasonable accuracy. However, to obtain all peculiar details of the KER, the exact electric field has to be used during the TDSE calculations.

\FloatBarrier

\section{Conclusion and outlook}\label{sec:conclusion}

The direct solution of the full TDSE for a molecular model interacting with short laser pulses of moderate intensity revealed that two well-known dissociation mechanisms, VED and BSD, are adequate to qualitatively understand the dissociation process. Already the simplest implementations of these mechanisms, first order perturbation theory for VED and adiabatic dynamics on a single Floquet PES for BSD, lead to a good quantitative agreement with the full TDSE in their respective regions of validity. The transition from VED to BSD, quite incomprehensible from the point of view of these very different mechanisms, proceeds rather unspectacular, but marks the onset of dynamical electron-nuclear correlations in the laser-driven molecular dynamics. 

For larger molecules with multi-dimensional Born-Oppenheimer PESs and multi-dimensional Floquet PESs, the resulting multi-dimensional structure of resonance frequencies might be to complicated for such a straight-forward discussion of VED and BSD as given in this work. Nevertheless, we fully expect our general findings also to hold for realistic molecular systems, especially dimers, as long as the laser spectrum fits the molecular structure. It is, for instance, an interesting question how the observed transition region between VED and BSD will be affected by a change (realistically an increase) of the nuclear mass.

A rigorous and explicit inclusion of the electric field of the laser turned out to be the key to the qualitative and quantitative validity of the VED model. Especially for the half-cycle pulses, it became apparent that a realistic modeling of the laser is of much greater importance as compared to conventional pulses of several ten femtoseconds duration. Still, the exemplary comparison of a half-cycle model pulse and a realistic attosecond pulse lead to good agreement for the (central)  electric field, the total dissociation probability, and, to certain extent, for the dissociation KER. In particular, nuclear motion during the temporal wings of the experimental attosecond pulse was found to be negligible, as concluded from the good agreement between VED model and exact result. 

However, this is not expected to hold in general. The aspiration for attosecond pulses with larger peak intensities results, probably inevitably, in higher intensities during the pulse wings. This will lead to non-negligible nuclear motion during these wings and thus much more involved dissociation dynamics.  Ultimately, it also opens possibilities to tune, control, and prepare chemical reactions by elaborately tailoring the electric field of attosecond (FWHM) pulses over their whole duration of femtoseconds, possibly granting access to electronic and nuclear time scales with a single laser  pulse. This poses a real challenge for laser physicists and also calls for advanced dissociation models applicable to rather large molecular systems and arbitrarily complicated electric fields.

\begin{acknowledgments}
We thank the authors of \cite{hassan_2016}, especially M.~Hassan and E.~Goulielmakis, for providing the data of the optical attosecond pulse. We gratefully acknowledge financial support from the Deutsche Forschungsgemeinschaft through the Normalverfahren (Nr. SCHM 957/10-1).
\end{acknowledgments}

\FloatBarrier

\end{document}